%% file: main.tex
\documentclass[runningheads]{llncs}
\usepackage[T1]{fontenc}
\usepackage{graphicx}
\usepackage{hyperref}
\usepackage{color}

\usepackage{amsmath} 
\usepackage{enumitem}
\usepackage{xspace}
\usepackage[table]{xcolor}
\usepackage{subcaption}
\usepackage[capitalise, noabbrev]{cleveref}
\usepackage{graphicx}
\usepackage{array}
\usepackage{booktabs}
\usepackage{float} %
\usepackage{comment}
\usepackage{csquotes}
\usepackage[normalem]{ulem}
\usepackage{wrapfig}
\usepackage{multirow}

\usepackage{color-edits}
\addauthor{cm}{black}

\newcommand{\code}[1]{\texttt{#1}\xspace}
\newcommand{\game}[1]{\texttt{#1}\xspace}
\newcommand{\prompt}[1]{\texttt{``#1''}\xspace}
\newcommand{\rplaycat}[1]{{\color{cyan!50!black}{\textsc{#1}}}\xspace}
\newcommand{\rplaycatTable}[1]{\textcolor{cyan!50!black}{\textsc{#1}}\xspace}

\newcommand{\rplayele}[1]{%
  {\color{cyan!30!black}\textit{#1}}\xspace}
  
\newcommand{\sysname}{\textsc{ImaginAItion}\xspace}
\newcommand{\paragraphBold}[1]{\paragraph{\emph{\textbf{#1}}}\xspace}

\definecolor{cquote}{HTML}{3c4043}

\newcommand{\supplement}{\href{https://github.com/mqo00/ImaginAItion/blob/main/study_material/supplements.pdf}{Supplements}}
\newcommand{\gitrepourl}{\url{https://github.com/mqo00/ImaginAItion}}

\begin{document}

\title{``GenAI Defaults to Bias!'' Gamify AI Literacy through Reflections on Prompts}

\titlerunning{\sysname: GenAI Literacy Game}
\author{
Qianou Ma\inst{1}\orcidID{0009-0002-8634-130X} \and
Megan Chai\inst{1} \and
Yike Tan\inst{1} \and
Jihun Choi\inst{1} \and
Jini Kim\inst{1} \and
Erik Harpstead\inst{1} \and
Geoff Kauffman\inst{1} \and
Tongshuang Wu\inst{1}\orcidID{0000-0003-1630-0588}
}

\authorrunning{Q. Ma et al.}

\institute{
Carnegie Mellon University, Pittsburgh, PA 15213, USA\\
\email{\{qianoum, mvchai, yiket, jihunc, jinik, eharpste, gfk, sherryw\}@cs.cmu.edu}
}

\maketitle              %
\begin{abstract}

As Generative AI (GenAI) becomes widespread, it is increasingly important for the public to understand the model's behaviors and biases.
However, existing AI literacy efforts miss opportunities to engage the general public to reflect on enduring GenAI bias and behaviors (e.g., how GenAI defaults to its internal bias in response to ambiguous or challenging prompts).
In this work, we introduce \sysname, a multiplayer game to help adults better reflect on GenAI bias and understand GenAI behaviors. 
\sysname is grounded in reflective play to surface GenAI limitations by encouraging players to manipulate prompt specificity (e.g., an underspecified prompt ``CEO'' defaults to a white man).  
From ten sessions (n=30), we find that the game significantly improved players' understanding of GenAI behaviors by 35\% in accuracy. Qualitative analysis showed how game mechanisms supported player reflections, including on prompting strategies to mitigate GenAI bias.
Our work demonstrates a viable pathway to scale GenAI literacy through playful, social interventions resilient to rapidly evolving technologies.

\end{abstract}
\input{sections/intro}

\input{sections/related}

\input{sections/design}

\input{sections/game_study}

\input{sections/results}

\input{sections/discussion}

\section{Conclusion}

\sysname is a party game designed to foster reflection on GenAI behaviors. 
In a study with 30 adults, we showed that reflective play could engage adults in exploring GenAI's biased defaults and prompting strategies to disambiguate human intents. By leveraging underspecification, peer discourse, and iterative experimentation, \sysname fostered reflection on enduring challenges of GenAI specification and significantly improved the accuracy of players' understandings over GenAI behaviors. Our analysis also offers design insights for future-resilient AI literacy interventions amid rapidly evolving technologies.

\begin{credits}
\subsubsection{\ackname} 
Thanks to all the participants for the study, and thanks to all LearnLab members who provided feedback on this work. Thanks to the National Science Foundation (award CNS-2213791 (SPLICE), 2414915) and Google Academic Research Award for partial support of this work.
\end{credits}

\bibliographystyle{splncs04}
\bibliography{refclean}

\end{document}

%% file: sections/intro.tex
\section{Introduction}
\label{sec:intro}

As Generative AI (GenAI) becomes embedded in more domains and everyday life, it is increasingly important for the public to understand model behaviors and bias.
Recent {AI literacy} frameworks similarly emphasize that AI competency demands the ability to \emph{critically reflect on} AI behaviors, such as how human input shapes AI output and reveals AI bias~\cite{long2020ai,Bozkurt2024-ed,Chiu2024-sz}.
For example, GenAI defaults to its training bias when given ambiguous (e.g., an underspecified prompt \prompt{CEO} defaults to a white man) or rare prompts (e.g., \prompt{horse riding astronaut} defaults to an astronaut riding a horse; \cref{tab:prompt-ex}). These GenAI behaviors are referred to as \textit{underspecification} \cite{Chenyang2025-qo} and \textit{LLM-hard specification} \cite{Ma2025-wf}, which have been a \cmedit{universal} and enduring challenge for humans to anticipate GenAI outputs and effectively interact with various GenAI~\cite{Simonen2025-vi}.
However, few existing AI literacy interventions provide support for users to reflect on such enduring GenAI behaviors~\cite{Chiu2024-sz}. %
Moreover, AI literacy work typically targets formal settings or narrow age groups, limiting broader public access~\cite{Casal-Otero2023-ll,Biagini2025-tv}.

Gamification offers a potential solution: it has emerged as a promising strategy for boosting AI literacy engagement across age groups including adults~\cite{Zhan2024-cn,ng2024fostering}, and frameworks on \textit{reflective play} \cite{Miller2024-wa} suggest games could be a medium for critical reflection. 
However, GenAI poses unique challenges as a subject of reflection.
Reflection fundamentally requires interpreting assumptions and observations~\cite{Mezirow1990-ka}, but GenAI's unstructured nature and vast output space make it hard for humans to form effective hypotheses or test assumptions, subsequently making human reflection on GenAI behaviors particularly difficult~\cite{Rapp2025-et,Di-Lodovico2025-ng}. 
These challenges prompt us to ask: \\
\textbf{Research Question: How to design games to support critical reflection by adults on GenAI behaviors, internal bias, and prompts?}

We explore this question by designing \sysname, a multiplayer party game that aims \cmedit{at a particular aspect of GenAI literacy:} fostering reflection on the enduring GenAI's behaviors of under- and challenging specification (\cref{subsec:design-rationale}) via manipulating prompts. 
\sysname effectively surfaces reflections by encouraging players to hypothesize about model behavior (e.g., ``Will it default to a white man if I don't specify demographic?'') and test assumptions through deliberate omissions in minimalist prompts. %
Inspired by party games~\cite{drawful} and the reflective play framework~\cite{Miller2024-wa}, \sysname engages players in an interactive loop of experimentation.

Through a playtest study with 30 adults across 10 trios (\cref{sec:game-study}), we analyze players' reflections on GenAI behaviors. Our analysis (\cref{res:default}) suggested that players' \textbf{reflections significantly calibrated their GenAI understanding pre-to-post gameplay} by 35\% in accuracy. Players also seemed to start reflecting on models' default bias, recognizing the social construction of bias, and forming prompting strategies to mitigate model biases.
These findings inform design directions for reflective AI literacy games (\cref{sec:discuss}), emphasizing deliberate constraints of the prompting freedom and coverage of enduring AI behaviors facing constant model updates.
In summary, we contribute:
\begin{itemize}[labelwidth=*,leftmargin=1.3em,align=left]
    \item \textbf{Empirical observations} of \textit{what} and \textit{how} players reflect on GenAI behaviors, internal bias, and prompts during reflective play (\cref{sec:results}); %
    \item \textbf{Design insights} for building future-resilient GenAI literacy interventions under unconstrained inputs and stochastic outputs of GenAI (\cref{subsec:fast-ai-design}); %
    \item \textbf{ImaginAItion} (\cref{fig:gameplay}), a multiplayer game that promotes reflective explorations \cmedit{on underspecification} for GenAI literacy (\cref{subsec:design-rationale}). 
    {[Code and supplemental study materials available at: \gitrepourl]}
\end{itemize}

\input{figures/gameplay}

%% file: figures/gameplay.tex
\begin{figure*}[t]
    \centering
    \begin{subfigure}[t]{0.49\textwidth}
        \includegraphics[width=\textwidth, clip, trim=15 15 15 75]{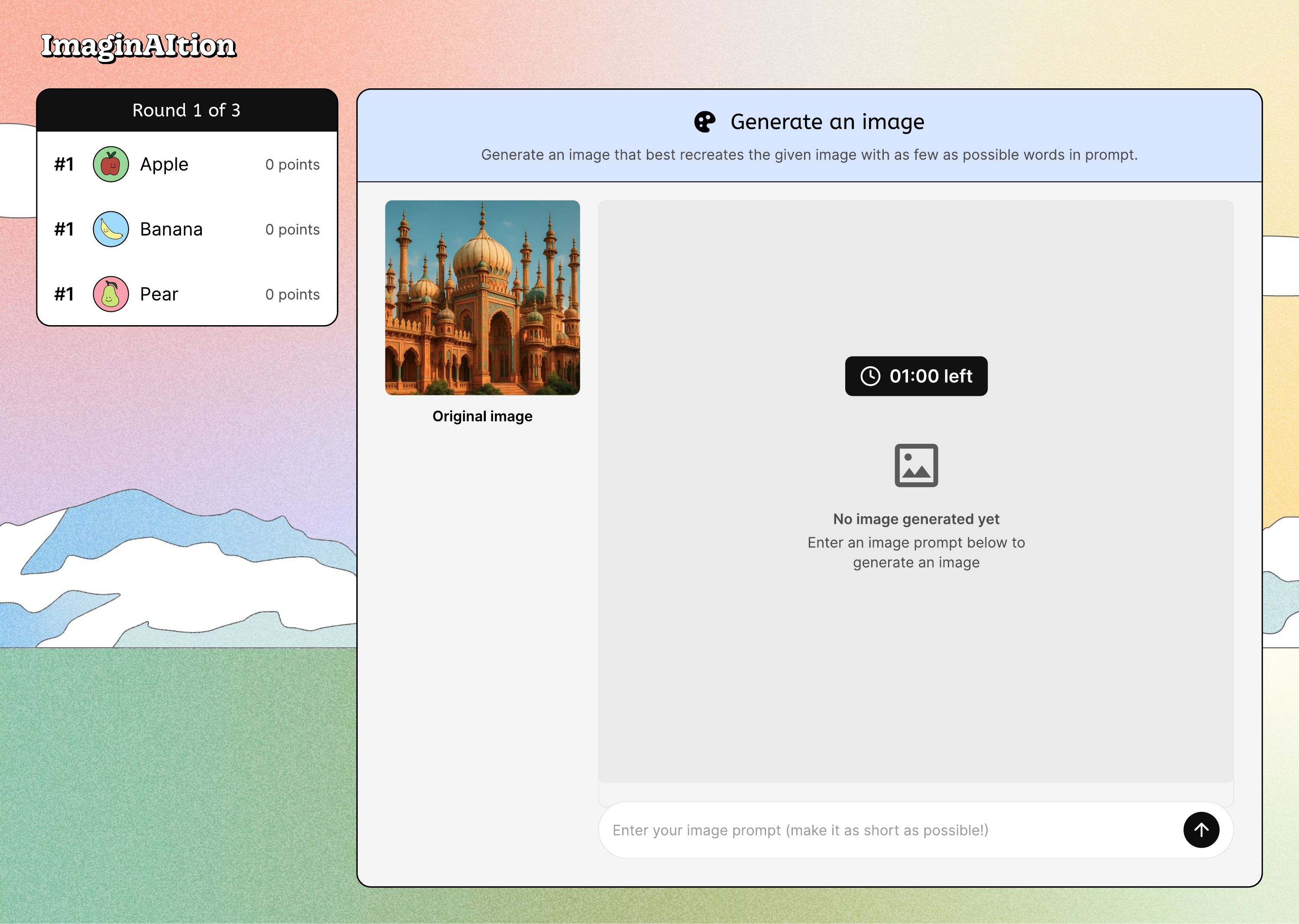}
        \vspace{-10pt}
        \caption{\textbf{\game{prompt-turn}} (70 seconds, excluding image generation): Given an image, creates a prompt to reproduce it as closely as possible, but using as few words as possible.}\label{fig:prompt}
    \end{subfigure}
    \hfill
    \begin{subfigure}[t]{0.49\textwidth}
        \includegraphics[width=\textwidth, clip, trim=15 15 15 75]{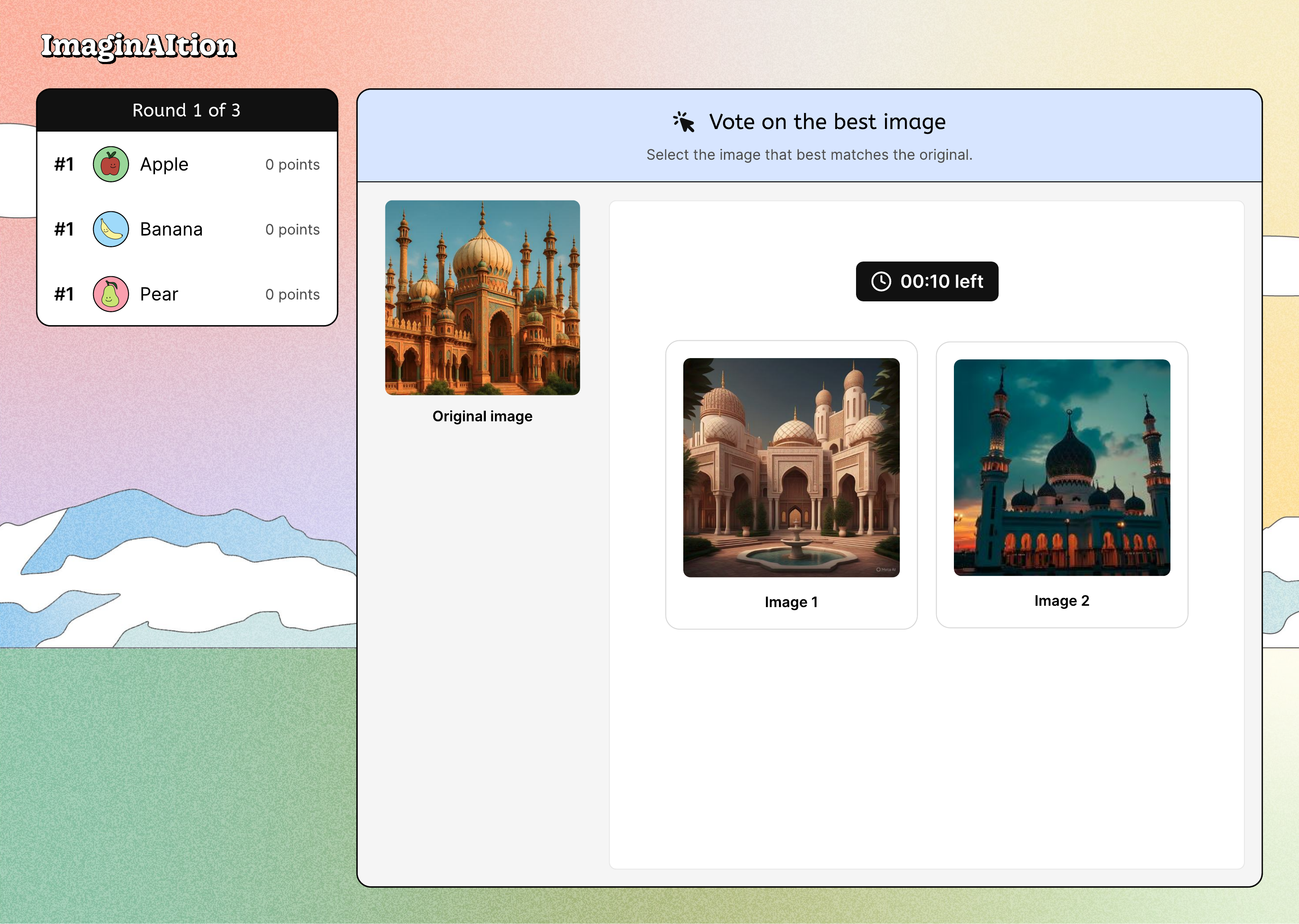}
        \vspace{-10pt}
        \caption{\textbf{\game{vote-turn}} (20s): Players review the generated images from others and vote on which is most similar to the original.}\label{fig:vote}
    \end{subfigure}
    
    \medskip
    \begin{subfigure}[t]{0.49\textwidth}
        \includegraphics[width=\textwidth, clip, trim=15 15 15 75]{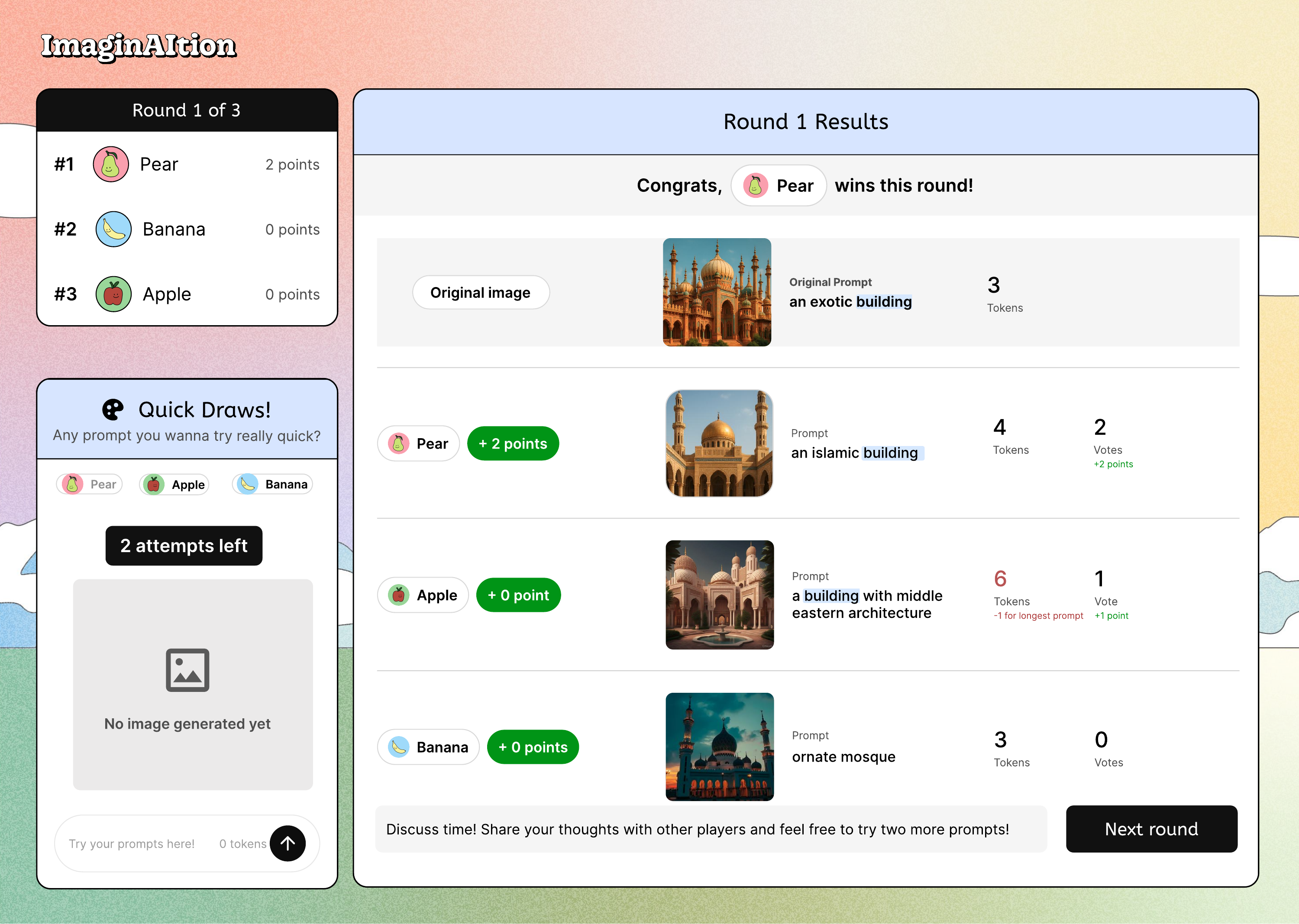}
        \vspace{-10pt}
        \caption{\textbf{\game{reveal-turn}} (untimed): Prompts, images, votes and scores are revealed. Reflection and discussion are encouraged with the Quick Draw chatbot, where players can test alternative prompts.}\label{fig:reveal}
    \end{subfigure}
    \hfill
    \begin{subfigure}[t]{0.49\textwidth}
        \includegraphics[width=\textwidth, clip, trim=15 60 15 30]{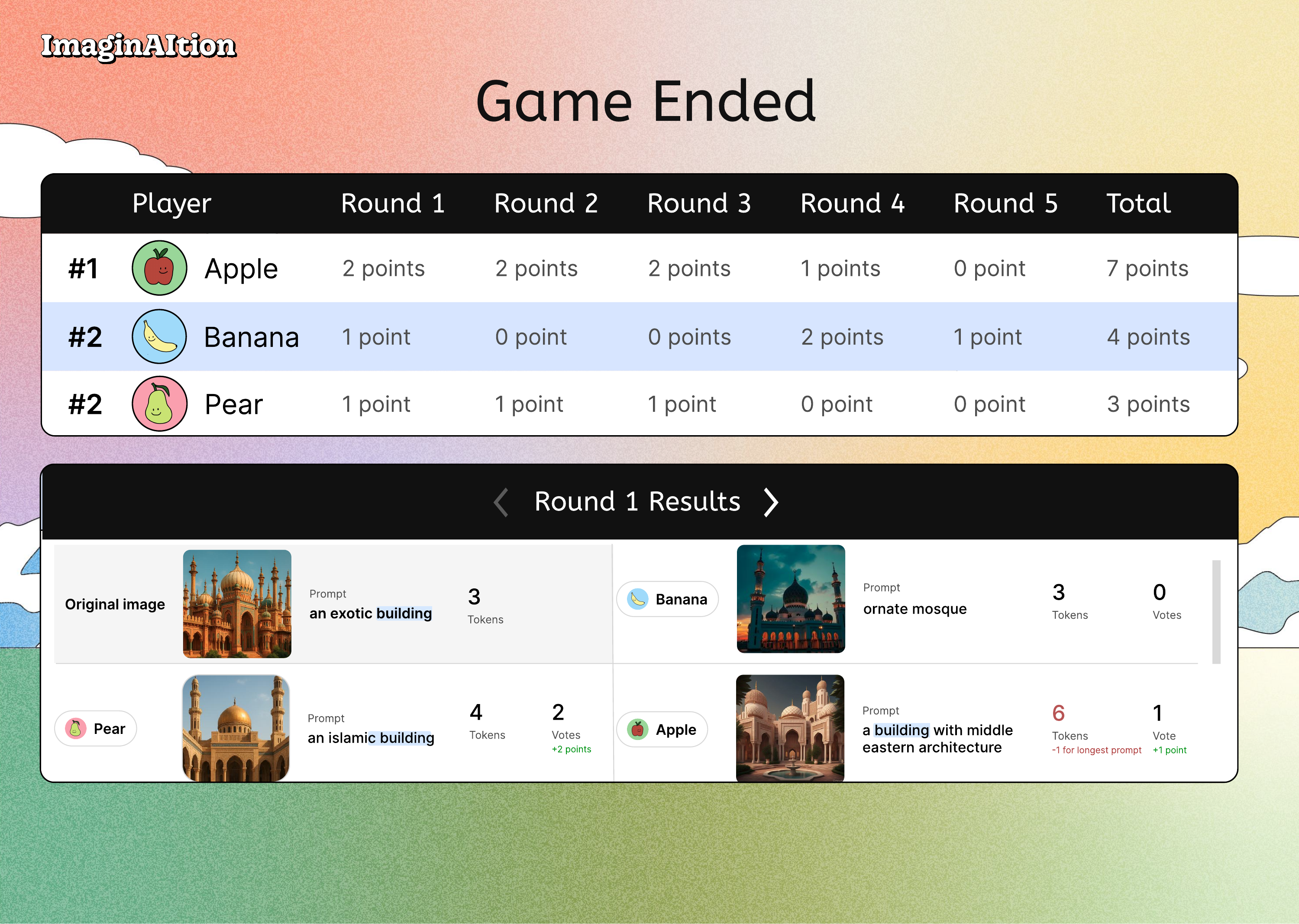}
        \vspace{-10pt}
        \caption{\textbf{\game{scoreboard}} (untimed): the final leaderboard with a summary of each round's results. Players earn one point for each vote. The longest prompt each round receives a one-point penalty.}\label{fig:scoreboard}
    \end{subfigure}
    \caption{An example sequence from \sysname gameplay.
    } \label{fig:gameplay}
\end{figure*}

%% file: sections/related.tex
\section{Related Works}
\label{sec:relate}

\paragraphBold{AI and GenAI Literacy and Instruction}
\label{subset:ai-literacy}

AI literacy refers to the ability to understand how AI operates, effectively utilize it, and critically evaluate the outputs of AI systems \cite{long2020ai}. AI literacy instruction for the general public remains underdeveloped \cite{Laupichler2022-fa}, as most initiatives have focused on K12 learners and resource-intensive formats like workshops and tutoring systems \cite{Casal-Otero2023-ll,Biagini2025-tv}. Recent work shows that adults (including college students and knowledge workers) suffer from overreliance on GenAI, which reduces critical evaluation and reflection~\cite{Melisa2025-tl,Lee2025-in}. Nonetheless, GenAI literacy requires critical reflection, especially on enduring and challenging behaviors inherent to generative models \cite{Chiu2024-sz}.  

Specifically, \textit{underspecification} in prompts, or the omission of requirements in prompts that make various outputs possible, can trigger GenAI default patterns that may not align with the user's intent~\cite{Chenyang2025-qo}. It is a \cmedit{universal behavior among different GenAI} that underspecified inputs can output biased representations caused by GenAI's training data~\cite{Simonen2025-vi}. LLM-hard or \textit{challenging specification} can also cause GenAI to produce unexpected outputs \cite{Ma2025-wf}. 
However, existing AI literacy efforts provide limited support for reflection on enduring GenAI behaviors of under- and challenging specification \cite{Chiu2024-sz,Lee2025-in}; this gap motivates our work.

\paragraphBold{Game-based Learning for AI Literacy}
\label{subsec:al-game}

Games are a promising medium for AI literacy due to their accessibility and ability to engage learners across age groups \cite{Zhan2024-cn}. Google's \emph{Quick, Draw!} \cite{quickdraw}, which adapts from popular party games like \emph{Pictionary} and \emph{Drawful} \cite{drawful}, demonstrates how playful interaction can surface AI capabilities and limitations at scale. 
Games for AI literacy have explored making complex AI topics more approachable \cite{ng2024fostering,Villareale2022-po}. However, these games only enable interactions with constrained AI systems, which are less useful for GenAI literacy where the models present \emph{open-ended inputs and unpredictable outputs}. Recent games for GenAI literacy have explored exposing bias and model limitations~\cite{Villareale2023-bm,Solyst2024-os}. However, these games only support \emph{limited interaction} with GenAI models, where players cannot actively shape model behaviors through their own inputs. As a result, players miss opportunities for critical reflection, which often emerges when learners test hypotheses and confront unexpected outputs~\cite{Miller2024-wa}. Additionally, some existing games target specific, short-lived GenAI behaviors that risk becoming outdated as GenAI systems rapidly evolve~\cite{Chenyang2025-qo}. We aim to address these gaps by designing a game that exposes \emph{persistent GenAI behaviors} and fosters \textit{critical reflection} with active interactions.

\paragraphBold{Reflection and Reflective Play}
\label{subsec:relate-reflect}

Games provide fertile ground for reflection and self-explanation~\cite{Richey2024-dm} by enabling safe spaces for experimentation, failure, and discovery \cite{Giere2003-ib}.
Reflection can be understood as cognitive or behavioral reappraisal that \cmedit{requires active interpretations of assumptions and observations, which could} lead to new insights or changes in judgment~\cite{Mezirow1990-ka}. Reflective play has been applied to personal, ethical, and societal issues, such as fostering healthy eating habits \cite{Miller2024-wa}. 
Villareale et al. incorporate reflection in games and expose how gameplay influences players' mental model development \cite{Villareale2022-po} or misconceptions of AI~\cite{Villareale2023-bm}. 
Recent work formalizes frameworks for reflective play \cite{Miller2024-wa}, including mechanisms such as cognitive dissonance, self-explanation, and social discourse, while also highlighting the difficulty of eliciting \textit{transformative, exo-game reflections}, \cmedit{which are reflections} that persist beyond game content. This challenge is particularly salient for GenAI literacy, where the goal is to support reflection on GenAI systems encountered in everyday life. Building on prior work, we apply reflective play principles and empirically examine how gameplay can elicit critical, socially situated reflection on enduring GenAI behaviors.

%% file: sections/design.tex
\section{The Design of \sysname}
\label{sec:design}

\paragraphBold{Reflection Goal (RG).}
\label{subsec:reflect-goals}

Observing the gap in GenAI literacy tools that promote reflection (\cref{sec:relate}), we create \sysname (\cref{fig:gameplay}), a lightweight game for adults to critically reflect on the core and enduring GenAI challenge:

\begin{enumerate}[labelwidth=*,leftmargin=2.6em,align=left,label=RG]%
\item \textbf{The under- and challenging- specification.} \label{rg:spec}
Players should reflect on the interplay between \emph{model defaults} and \emph{user specification}. 
Players should recognize the model's ``mental shortcuts'' and reflect on how and why it fills gaps under uncertainty or challenging constraints.
\end{enumerate}

Pre-training effectively encodes a data-driven ``world model'' within GenAI, causing under-specified or unusual (challenging specification) prompts to reveal learned defaults~\cite{Simonen2025-vi,Chenyang2025-qo}. 
For example (\cref{tab:prompt-ex}), \prompt{CEO} without demographics context generated a stereotypical default of a white man, \prompt{A pretty cow} without stylistic cues (e.g., \prompt{cartoon style}) yields realistic imagery. Some defaults resist even explicit instructions (e.g., \prompt{a horse riding an astronaut} still produces \emph{an astronaut riding a horse} instead).
\label{subsec:design-promptex}
To support our reflection goal in game, we adopt from the existing GenAI bias taxonomy~\cite{Vazquez2024-le} and reported examples~\cite{Simonen2025-vi} to purposefully select prompts like above, which reveal different GenAI behaviors when prompts are underspecified or include challenging specifications.

\label{subsec:design-overview}
\label{subsec:design-rationale}

\input{figures/table_goal_alignment}

\paragraphBold{Game Overview.} 
We conducted 10+ iterations of prototyping \cite{Ma2025-hi} in order to instantiate the reflective play framework~\cite{Miller2024-wa}.
We design \sysname as a multiplayer web-based game, with reflective play mechanisms to support the above reflection goal. %
Taking inspiration from Drawful~\cite{drawful}, \sysname asks all players to compete in \emph{recreating a target image using the shortest possible prompt}.
\cref{fig:gameplay} shows a complete sequence of three core turns per round: 

\textbf{Prompting}, where players each form their hypothesis on what details can be omitted to achieve close image reproduction with few words;
\textbf{Voting}, where players vote for the closest image reproduction; 
\textbf{Revealing}, where players discuss and hypothesize GenAI behaviors based on the revealed prompts, images, and their corresponding votes and scores, and test their hypotheses by retrying some prompts in a \textbf{Quick Draw} chatbot panel.
Their final scores are revealed after six rounds, accumulating the number of votes they received, minus the penalty for the longest prompt.

\paragraphBold{Mechanism Design.}
As shown in \cref{tab:rg-game-alignment}, we design \sysname to instantiate design patterns in the reflective play theory~\cite{Miller2024-wa}: \emph{Disruptions} (challenge assumptions), \emph{Slowdowns} (create space to reflect), \emph{Questioning} (provoke critical thought), \emph{Revisiting} (re-experience past choices), and \emph{Enhancers} (extend reflection beyond the game):

\input{figures/table_prompt_simplified}

\textbf{\game{Constrained prompt} as hypothesis anchoring.}
\sysname asks players to replicate a target image with the \textit{shortest} prompt, penalizing the longest prompt.
This brevity incentive pushes players into under-specification (\ref{rg:spec}), revealing what they assume the model will ``fill in'' by default --- players' omissions in the prompt encode their mental models of the GenAI.
This mechanism not only facilitates reflection by \rplaycat{Questioning} (as players need to \rplayele{self-explain} what they put in prompts), but it also acts as the anchor for downstream \rplaycat{Disruption} and \rplaycat{Revisit}.
It sets up comparison across prompts and outcomes, priming players to reflect on when and why the model fails to behave as expected.

\textbf{\game{Structured contrast} to surface \rplaycat{Disruption} and prioritize \rplaycat{Questioning}.}
\sysname supports reflection through multiple layers of contrast.
Players compare their own prompts to their generated images, and then compare both to peers' outputs or the \game{original-prompt}, observing how their best efforts of guessing model behavior may be sub-optimal and receive fewer \game{votes}. 
These individual and in-group outcome mismatches produce \rplaycat{Disruption} where expectations are not met, encouraging players to \rplaycat{question} their \rplayele{self-explanation} on GenAI behaviors and test new \rplayele{hypothesis} in \game{quick-draw}.
The image generation delay creates a \rplaycat{slowdown} that encourages forming concrete expectations, while the untimed \game{reveal-turn} supports \rplayele{lingering defeat} and richer discussion.
These contrast mechanisms expose GenAI behaviors under different levels of specificity in prompts (\ref{rg:spec}).

\textbf{\game{Multi-party review} for calibration and perspective shift.}
The multiplayer structure \rplaycat{enhances} reflection.
Seeing how other players \game{vote} and what they chose to specify (or omit) reveals differences in perceptual priorities (e.g., color over shape, or character over context), expectations, and prompting strategies.
Through \rplayele{social discourse} and the \rplayele{explicit reflection prompt}, players identify blind spots and collectively develop a broader hypothesis space for how GenAI react to prompt specificity (\ref{rg:spec}).

%% file: figures/table_goal_alignment.tex
\begin{table*}[t]
\setlength{\parindent}{0pt}

\centering
\fontsize{7.3}{8.3}\selectfont
\caption{Mapping \sysname mechanism designs to Reflective Play \cite{Miller2024-wa}.}
\vspace{5pt}
\begin{tabular}{p{0.21\textwidth} | p{0.78\textwidth}}

\toprule
\textbf{Mechanism} & \textbf{Reflective Play theory mapping} \\
\midrule

\textbf{\game{Constrained prompt} }
    \newline \emph{in} \game{prompt}, \game{score}
    & \rplaycatTable{Questioning} $\to$ \rplayele{Self-Explanation}: players commit assumptions about what the model will ``fill in'' when anticipating model behaviors to their short prompt. \\

\midrule

\textbf{\game{Structured contrast}}
    \newline \emph{in} \game{prompt}, \game{vote}, \newline\game{reveal}
    & 
        \rplaycatTable{Disruptions} $\to$ \rplayele{Narrative Twist}: expectation–output mismatches create cognitive conflict; \newline
        \rplaycat{Disruptions} $\to$ \rplayele{Confrontation}: cross-comparisons expose suboptimal or misaligned choices; \newline
        \rplaycat{Slowdowns} $\to$ \rplayele{Weighting Mechanics}: generation delays sharpen expectations; \newline
        \rplaycat{Slowdowns} $\to$ \rplayele{Lingering Defeat}: untimed review supports analyzing failure before retry; \newline
        \rplaycat{Questioning} $\to$ \rplayele{Hypothesis Testing}: \game{quick-draw} enables targeted prompt edits for probing. 
\\
\midrule

\textbf{\game{Multi-player review}}
    \newline \emph{in} \game{vote}, \game{reveal}
    & 
        \rplaycatTable{Enhancers} $\to$ \rplayele{Social Discourse}: peers' prompts/images surface blind spots; \newline %
        \rplaycat{Enhancers} $\to$ \rplayele{Explicit Encouragement}: prompt for discussion that support transformative reflection. 
\\ 
\midrule

\textbf{\game{Repeated exposure}}
    \newline \emph{in} \game{multi-round}, \newline\game{reveal}
    & 
        \rplaycatTable{Revisiting} $\to$ \rplayele{Killcam}: looking back at prompt failures supports reasoning and improvement; \newline
        \rplaycat{Revisiting} $\to$ \rplayele{Reflective Revisiting}: cumulative results reveal persistent patterns. \\
\bottomrule
\end{tabular}
\label{tab:rg-game-alignment}
\vspace{-5pt}
\end{table*}

%% file: figures/table_prompt_simplified.tex
\newcommand{\imgsize}{1.2cm}

\begin{table*}[tb]
\renewcommand{\arraystretch}{0.9}

\centering
\fontsize{8}{9}\selectfont
\caption{Example GenAI's behaviors for \ref{rg:spec}. The full table is in \supplement.} %
\vspace{5pt}
\label{tab:prompt-ex}

\begin{tabular}{
    >{\raggedright\arraybackslash}m{0.13\textwidth}  %
    >{\raggedright\arraybackslash}m{0.22\textwidth}  %
    >{\raggedright\arraybackslash}m{0.2\textwidth}  %
    >{\raggedright\arraybackslash}m{0.32\textwidth}  %
    >{\centering\arraybackslash}m{0.08\textwidth}    %
}
\toprule
\textbf{Category} & 
\textbf{Under-/Challenging specs} & 
\textbf{Default behavior} & 
\textbf{Example prompt and behavior} & 
\textbf{Ex. image} \\
\midrule

\textbf{Bias (Biological)} & 
Underspecify biological factors like age, sex/gender, disability & 
Defaults to historical gender roles or stereotypes &  
\prompt{CEO} defaults to middle-aged white male; Note that \prompt{CEO} demonstrates both biological and demographic bias. %
& 
\includegraphics[trim=20 400 50 0,width=\imgsize,height=\imgsize,clip]{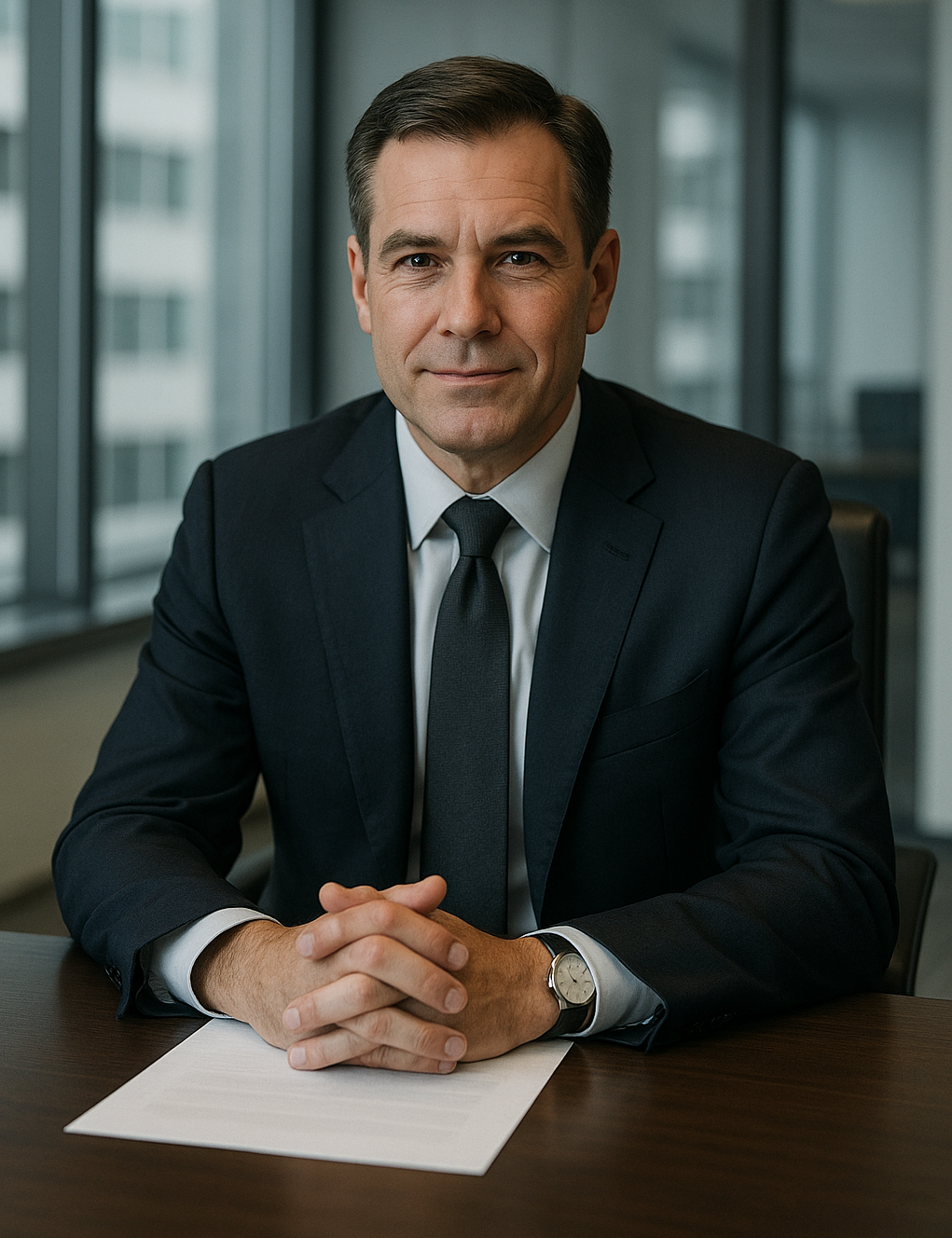} \\
\midrule

\textbf{Realistic style} & 
Underspecify style and uncommon request of abstract concepts / adjectives & 
Defaults to photo-realism & 
\prompt{A pretty cow} tends show abstract or anthropomorphic concepts such as ``pretty'' and ``sad'' in realistic styles and can be hard to interpret &
\includegraphics[trim=0 300 0 200, width=\imgsize,height=\imgsize,clip]{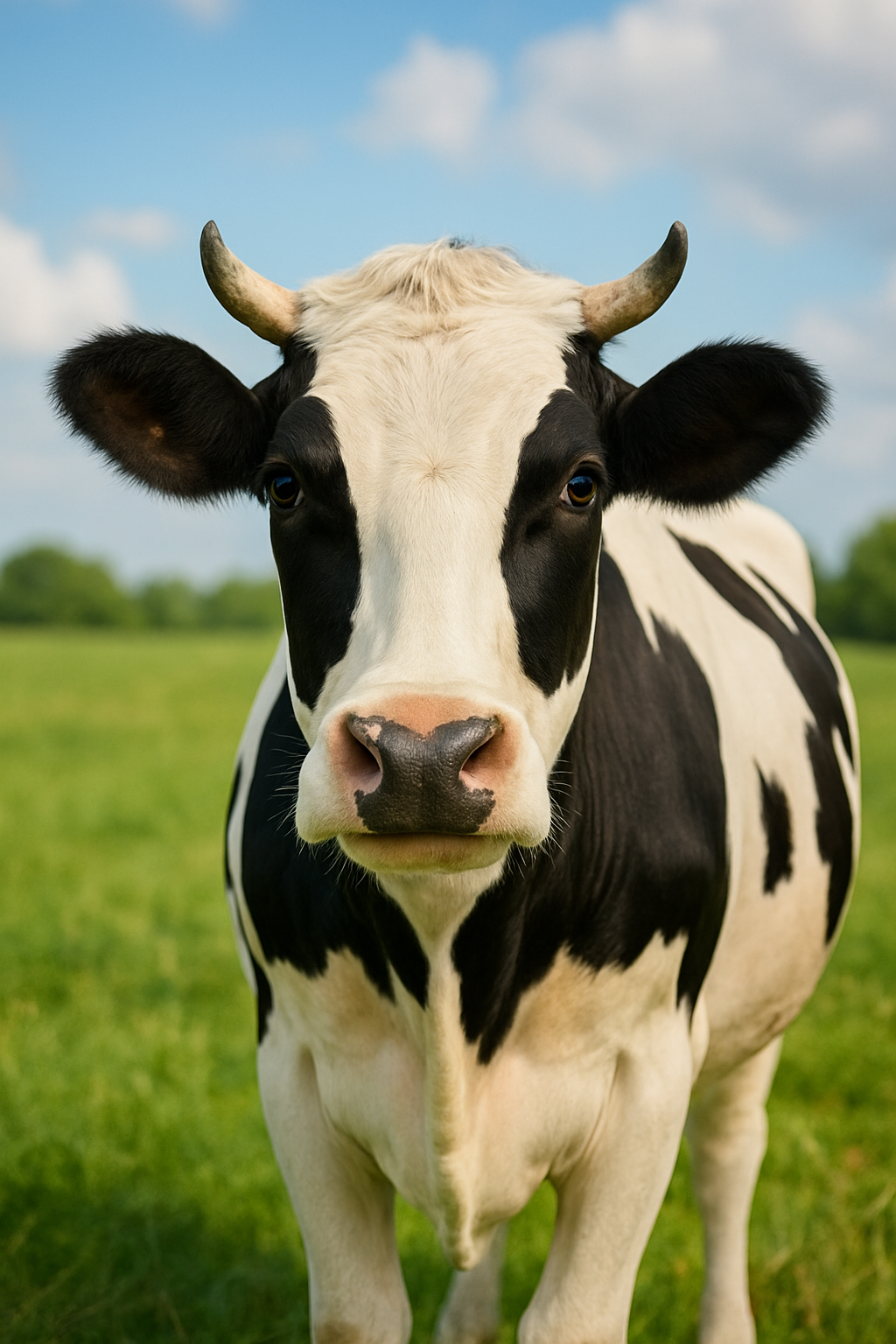} \\
\midrule

\textbf{Common co-occurrence} & 
Uncommon prompt reliant on negation or syntax parsing & 
Autocorrects to frequent patterns & 
\prompt{A horse riding an astronaut} portrays an astronaut riding the horse instead 
& 
\includegraphics[trim=0 0 0 0, width=\imgsize,height=\imgsize,clip]{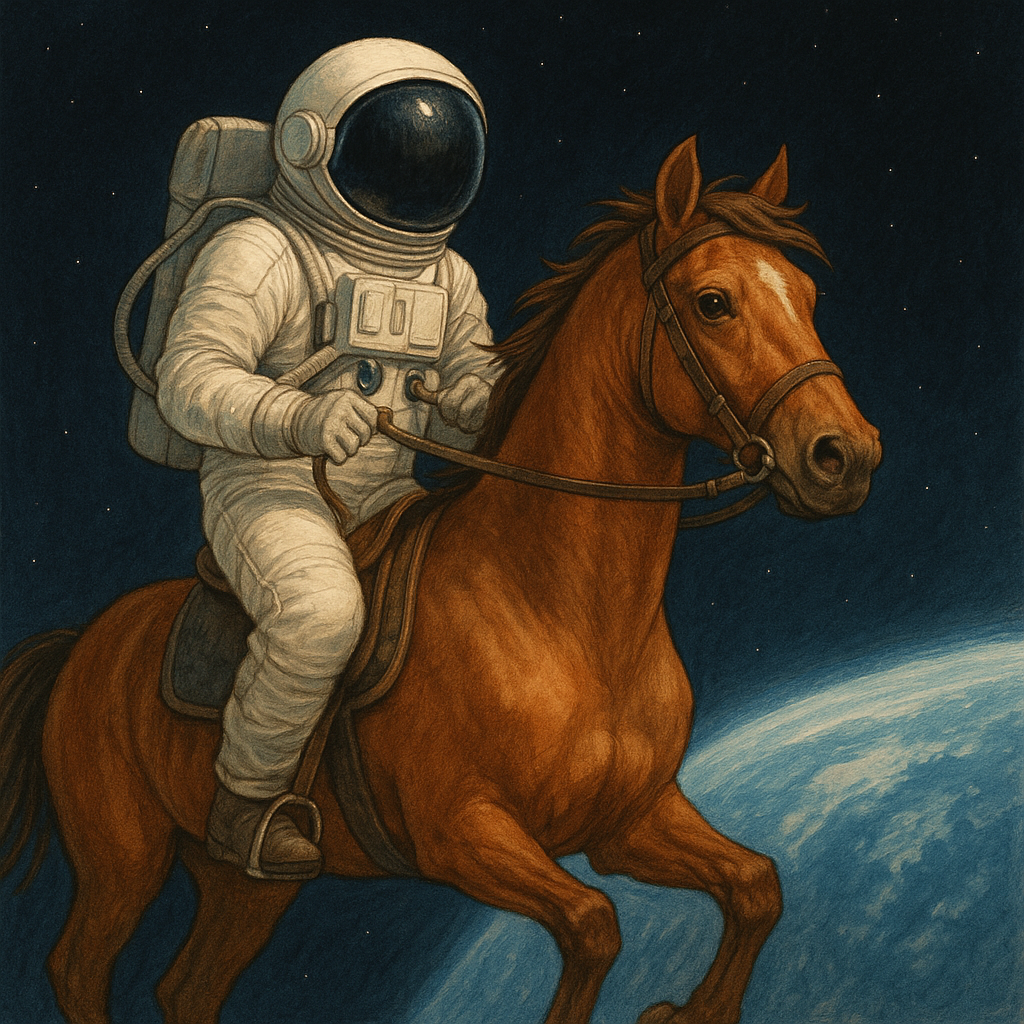} \\

\bottomrule
\end{tabular}
\vspace{-10pt}
\end{table*}

%% file: sections/game_study.tex
\section{Game Study}
\label{sec:game-study}

We conducted 60-minute online gameplay sessions with 10 groups of three (n=30) to examine \emph{what} and \emph{how} players reflect on GenAI behaviors related to our reflection goal (\cref{subsec:reflect-goals}), how players' understanding of GenAI behaviors shifted, and how \sysname's design choices shape these reflections. 
Participants were compensated with a \$15 Amazon gift card. 

Each session consisted of three phases: (1) a pre-survey collecting demographics, prior experience with GenAI, and baseline understandings of GenAI behaviors aligned with our reflection goal; (2) a tutorial followed by six rounds of gameplay (7 minutes each) covering different prompt categories (e.g., bias, realism, co-occurrence; selection criteria discussed in \cref{subsec:design-promptex}); and (3) a post-survey and a brief semi-structured group interview probing participants' reflections, learning, and gameplay experience.

\paragraphBold{Participants}

We recruited 30 adults (ages 18–53, $25.4 \pm 6.2$; 14 women) with varied self-rated familiarity with GenAI ($4.6 \pm 1.7$ on a 7-point scale). Thirteen participants self-identified as novices\footnote{How familiar are you with text-to-image GenAI tools (1-7)? Novice: score $\leq 4$. \label{footnote:expertise}}, providing natural variation in prior knowledge. To approximate authentic party game contexts, we primarily recruited friend groups, adding individual participants if needed to form 10 trios. \cmedit{Each participant is referred to as Gx-Py, where $x \in [1,10]$ denotes the group and $y \in [1,3]$ denotes the player within group}. Group compositions varied in friendship status, demographics, and GenAI experience, supporting diverse collaborative reflection (details in \supplement).

\input{figures/table_codebook_simplified}

\paragraphBold{Method}

Our analysis combined a lightweight pre-post assessment and thematic analysis to provide preliminary evidence of reflection and learning. Each participant answered \cmedit{seven open-ended questions each in pre- and post-survey}, probing their current understandings of GenAI behaviors in different categories (e.g., “How does GenAI represent social and cultural elements in images?”\footnote{The questions were designed and iterated to align with prompt categories used in the game, like bias and realism (\cref{tab:prompt-ex}). Full questions are provided in \supplement. We also include questions about GenAI’s ability to render text and body parts as a baseline to check whether players have up-to-date understandings for GenAI (as the state-of-the-art models no longer produce garbled text or distorted body parts~\cite{Borji2023-tp}).}. 

Quantitatively, responses are graded using a simple rubric: \textit{accurate} (1), \textit{no clear understanding} (0), or \textit{flawed} (-1), resulting in $3\times3=9$ combinations of understanding shift pre- to post-game (\cref{fig:code_distr}a; \cmedit{e.g., Enlightened (+1) = \textit{no} $\rightarrow$ \textit{accurate} understanding}, more examples in \cref{tab:codebook-ex}).
Two authors as domain experts independently annotated a 10\% random subset and achieved moderate to strong agreement (Cohen's $\kappa = 0.79$) \cite{McHugh2012-sc}; disagreements were resolved through discussion, and one author then graded the rest of the responses. 

Qualitatively, gameplay logs, interviews, and open-ended survey responses were thematically analyzed. Three authors adopted an open coding process and iteratively refined themes of reflection~\cite{fereday2006demonstrating}, including model fill-in-the-blank with bias, bias as a ubiquitous concept, and prompting strategies to mitigate bias.

%% file: figures/table_codebook_simplified.tex
\begin{table*}[tb]
\centering
\caption{Example of understanding shifts from pre- to post- gameplay. {See \supplement \ for full rubrics.}} %
\label{tab:codebook-ex}
\small
\begin{tabular}{
    p{0.22\linewidth}
    p{0.46\linewidth}
    p{0.3\linewidth}
}
\toprule
\textbf{Rubric} & \textbf{Description} & \textbf{Example(s)} \\
\midrule

\code{Enlightened} (+1): \newline No (0) $\rightarrow$\newline Accurate (+1) & Often seen in novices who start recognizing bias patterns or identifying capabilities and limits of GenAI, starting from don't know or irrelevant answer. & \textbf{Cultural (G3-P2):} “I don't know” $\rightarrow$ “leaning to the western culture and perceptions.” \\
\midrule

\code{Aligned} (+2): \newline Flawed (-1) $\rightarrow$\newline Accurate (+1) & Sweeping claims become nuanced, conditional statements; outdated views are updated to SOTA capacity; or vague generalizations become specific. & 
\textbf{Number \& Spatial (G6-P3):} “well” $\rightarrow$ “one needs to be very specific with positions of objects”\\

\midrule

\code{Misaligned} (-1): \newline No (0) $\rightarrow$ \newline Flawed (-1) & Often stems from overlooking or misattributing model limitation (e.g., expecting the model to infer unspecified text). & \textbf{Number \& Spatial (G6-P2):} “I don't know” $\rightarrow$ “really well.” \\

\bottomrule
\end{tabular}
\end{table*}

%% file: sections/results.tex
\section{Game Study Results}
\label{sec:results}

\subsubsection{Players understand GenAI behaviors significantly more accurately after gameplay, and self-rated novices benefit more.} 
\label{res:default}

\label{subres:positive-ro}

\input{figures/fig_code_distr}

A two-tailed paired $t$-test showed a significant increase in understanding scores from pre- to post-gameplay ($p < 0.001$, \cmedit{Cohen's $d = 1.25$, a large effect size \cite{lakens2013effectsize}})\footnote{\cmedit{As a conservative check, we aggregated scores at the group level, and a paired t-test on group means (N = 10) remained statistically significant ($p < 0.001$).}}. Accuracy ($N_\text{Accurate} / (N_\text{Accurate} + N_\text{No} + N_\text{Flawed})$) on GenAI understandings improved from 38.4\% to 73.0\%\footnote{\label{footnote:normalized_gain}\cmedit{To account for ceiling effect, we computed a normalized gain in understanding scores for each participant $i$ as $\frac{post_i - pre_i}{max - pre_i}$, where the $pre_i$ and $post_i$ are summed across seven survey answers (each graded as accurate = 1, unclear = 0, or flawed = $-1$). Two self-rated experts with $pre_i = max = 7$ were excluded, and the mean normalized gain was $M=0.44$.}}. These statistics offer preliminary quantitative evidence that reflective gameplay could be associated with improved understanding of GenAI behaviors, \cmedit{and we discuss limitations of our methods and designs in \cref{subsec:limit}.}

As shown in \cref{fig:code_distr}, 73.8\% of understanding on GenAI behaviors landed in \emph{accurate} states, as most participants either retained a correct understanding (\code{Confirmed}) or shifted toward a more nuanced one (\code{Aligned} or \code{Enlightened}). 
When directly asked whether the game helped improve their understanding of text-to-image GenAI's behaviors during the post-survey (1 = not helpful at all, 7 = extremely helpful), participants rated the game as highly effective ($5.7 \pm 0.9$). 
And their self-rated confidence in understanding GenAI's behaviors (1 = not at all confident, 7 = extremely confident) also increased by about one point from the pre-survey ($4.2 \pm 1.6$) to the post-survey ($5.1 \pm 1.2$).

\begin{wrapfigure}{tr}{0.45\textwidth}
\vspace{-15pt}
\includegraphics[trim=2cm 0 0 0, width=\linewidth,clip]{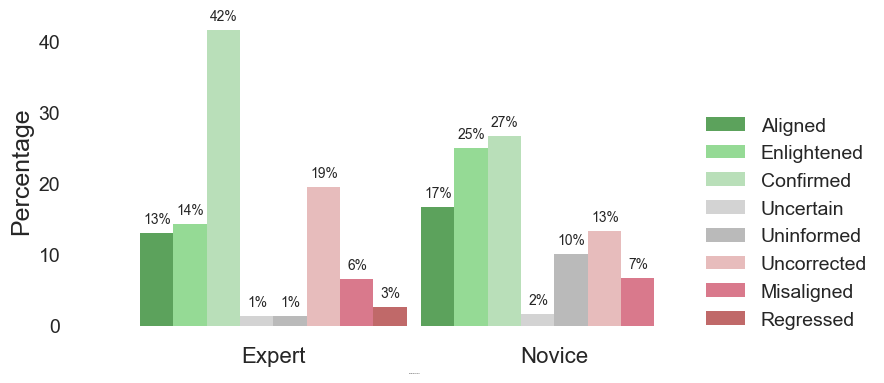}
\vspace{-20pt}
\caption{Reflection outcomes distribution by self-rated expertise.}
\vspace{-15pt}
\label{subfig:code-expert}
\end{wrapfigure}

\label{subres:expert-vs-novice}
We see more nuances of understanding shifts when broken down by self-rated expertise$^\text{\ref{footnote:expertise}}$. Accuracy on GenAI understandings for self-rated novices improved from 27.8\% to 69.6\%, while self-rated experts had a better accuracy pre-game (46.2\%) but achieved a smaller accuracy gain (29.2\% < 41.8\%)\footnote{\cmedit{Both self-rated novices and experts' increase in pre- to post-game understanding scores are significant (paired t-tests; $p < 0.001$). Self-rated experts achieved a mean normalized gain (\cref{footnote:normalized_gain}) of $M = 0.42$, and novices $M = 0.48 $.}}.
As shown in \cref{subfig:code-expert}, more \code{Uncorrected} (19\% > 13\%) understanding suggested that \textbf{self-rated experts may have firmer beliefs that are harder to challenge}. 
In comparison, self-rated novices showed more \code{Enlightened} (25\% > 14\%), as they may gain new insights from the examples we provided in the game. %
However, the game \textbf{may fail to inform novices} (\code{Uninformed} 10\% > 1\%). Both experts (6\%) and novices (7\%) also risk becoming \code{Misaligned}, failing to observe calibrated patterns when they begin with \emph{no} understanding of GenAI behavior.

\subsubsection{Players reflect on how model defaults to fill-in-the-blank with bias.}
An episode of reflection happened during \game{quick-draw} for the round with original prompt \prompt{holding baby} (\cref{fig:G7-quickdraw-mother}). After seeing the very brief original prompt, players were inspired to try as short as possible prompts, which further exposed model default bias, and reflections carried over into post-game discussions: 

\begin{figure*}[t]
\includegraphics[trim=1cm 3.5cm 1cm 3cm, width=.9\linewidth,clip]{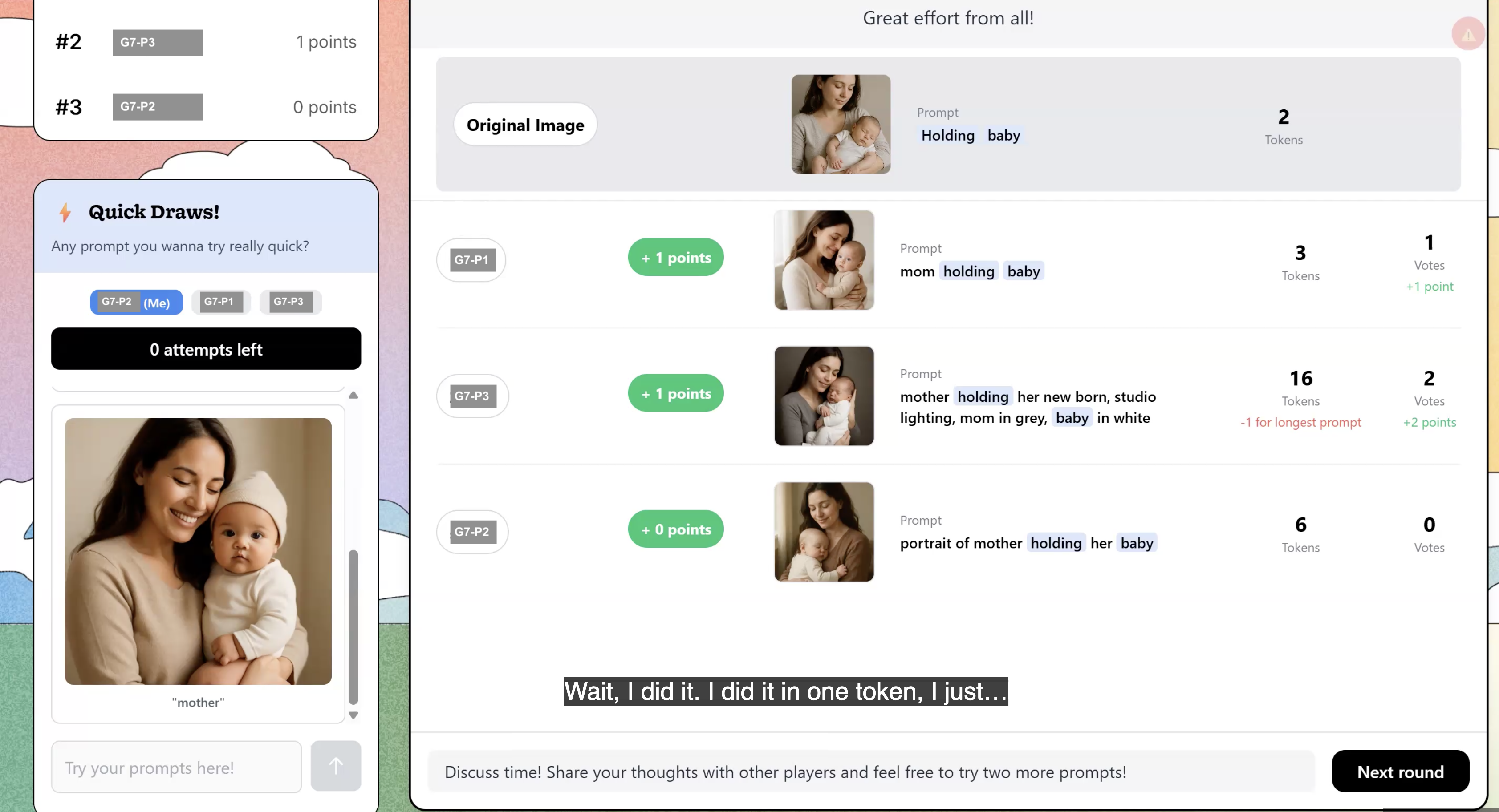}
\caption{G7-P2's \game{quick-draw} experiment to reduce the prompt length using just \prompt{mother}. The reproduced image is considered more similar to the original image than her prompt \prompt{portrait of mother holding her baby}.
}
\label{fig:G7-quickdraw-mother}
\end{figure*}

\blockquote{
G7-P2: I did it in one token --- I just wrote \emph{mother}. And it's pretty similar ... better than whatever my portrait of mother holding her baby came up with. I guess \emph{mother} is assumed to be you're holding your baby. \\
G7-P3: The social role of a mother is relative to the offspring that she produced. What? \\
(later in post-game interview)
G7-P1: I was also surprised — when G7-P2 prompted just \emph{mother}, the baby also appeared. \\
G7-P2: And it didn’t think of any other stage of child — not a toddler, middle schooler, or high schooler. It was just baby. GenAI's \emph{defaulted}. \\
G7-P1: I wonder if you prompt \emph{father} if a baby would appear. I would guess no, I feel like… it's going based off of what’s stereotypical, unfortunately. It would probably only put the baby with the mom. \\
G7-P3: Yeah, not very common either online or in real life.}

Our game's \game{structured contrast} seemed to bring \rplaycat{Disruption} and \rplaycat{Slowdown} to facilitate reflection.
G7-P1 described the wait time during image generation as a ``suspenseful moment'' and is fun ``like gambling'' when the generated images go against players' expectations. %
We observed a lot of playful ``aha'' and ``ohno'' moments of cognitive dissonance.
For example, when G10-P3 entered the prompt of \prompt{A white cis male with a brown beard smiling with a blue t shirt in a portrait picture} and the original prompt was revealed to be simply \prompt{A man}, he reacted with surprise: \textquote[]{Oh my god, it's just \emph{a man}!}

Some players tried to guess the AI's default, which further prompted them to reflect on \textbf{their \emph{own} perceptions on the AI biases.}
For example, G0-P3 reflected, \textquote{The moment where I thought models will give me a middle-aged White male CEO as default to CEO is insightful to me, it helps me reflect on my own implicit biases.} 

\subsubsection{Players reflect on ubiquitous bias in both models and humans.}
The \game{multi-round} gameplay and different prompt categories' repeated exposure might help exemplify that the model embeds systematic biases.
In the pre-survey, G5-P1 described that he did not know the mechanism of how GenAI depicts sociocultural elements in images and did not mention potential bias. In the post-survey, G5-P1 commented that there are \textquote[]{many stereotypical things, such as race, skin color, clothing, and even gender}. \sysname also seemed to {trigger \textit{transformative reflections} that resonate with players even outside of the gameplay}:
\blockquote[G5-P1]{GenAI is so problematic. But then I wonder --- isn't this precisely a reflection of human ways of thinking? For example, if we want to express something “Chinese,” wouldn’t a human painter also draw a Chinese knot? GenAI is merely replicating the flaws of human thinking. After all, culture itself is a product of social construction.}

Access to peer comparison seems to help players reflect on the \textbf{source of the bias} --- a participant in the pilot acknowledged how their own background influenced their perception directly: \textquote[G0-P1]{I was clearly biased when I saw the white coat, I immediately thought of my dad, then thought of a doctor, and I completely missed the rocket part. But you thought the rocket was more prominent -- that's your bias.}
Group diversity helped players conceptualize \textbf{\emph{bias} as a ubiquitous concept}: \textquote[G0-P2]{Models, and also humans, have stereotypes.} %
\textbf{Diversity also seemed to amplify the effectiveness of reflective play}. For example, we observed women in co-ed groups questioned gendered representations, such as when G10-P2 and G2-P2 experimented with ``a woman'' after seeing the \prompt{A man} prompt, or when G7-P1 tested whether prompting ``father'' also generated a baby (\cref{fig:G7-quickdraw-mother}). 
In comparison, more homogeneous groups might miss opportunities to interrogate model bias; for instance, all G8 players (three South Asian men) did not remark on the White male default in \prompt{A man}.

\label{subres:human-ai-diff}

\subsubsection{Players reflect on prompting strategies to mitigate bias and misalignment.}
\label{subres:prompt}
Our \game{constrained prompting} mechanism
might help players become
{more self-aware of their own and the AI's \emph{salience models}} on what \emph{must} be said --- \textquote[G5-P2]{What I think is important isn't in the original prompt}.
\textbf{When users fail to anticipate internal model bias, they may also fail to effectively control the output}.
For example, misalignment was observed when G5-P1 tried to reproduce \prompt{A man} using \prompt{David} and later \prompt{David selfie}, which turned to Michelangelo's sculpture (the later even holds a phone in hand).

We observed players reflecting on their prompting strategies for approaching model underspecification and misalignment. 
As we asked about lessons learned from the game in the post-survey, 20 out of 30 participants talked about prompting, and four strategies were most prominent: %

\begin{itemize}[labelwidth=*,leftmargin=1.3em,align=left]
    \item \emph{Be specific} to override defaults and explicitly specify necessary details. E.g., \textquote[G2-P1]{I'm justified in using a greater degree of specificity in my prompts. My mental model of which details to make explicit is also changing.} %
    \item \emph{Gamble} with concise prompts to exploit model defaults and biases. 
    E.g., \textquote[G2-P2]{There are moments when it's good to be more specific vs not -- you can play to AI's biases.}
    \item \emph{Iterate} with multiple attempts to converge on the desired result. 
    E.g., \textquote[G1-P1]{You probably can't avoid iterative prompting to get what you want.}
    \item \emph{Contrast} prompts to isolate discriminative features and test assumptions. 
    E.g., \textquote[G6-P2]{I understood how different parameters (amount of descriptions, details, model default behaviors) affected outputs with different prompts.}
\end{itemize}

Players also self-rated to be more confident in their ability to craft useful prompts: their perceived ease of creating effective prompts improved by about one point from pre-survey ($4.0 \pm 1.5$) to post-survey ($4.9 \pm 1.4$) on a 7-point scale (7 = extremely easy). As G7-P1 remarked, \textquote[]{I have a better understanding now of how generative AI works! I think that if I were to use generative AI to create images after playing this game, I would be able to generate images that more closely match my intended goals.} Follow-up studies can track if players' prompting strategies actually changed in real life after gameplay (\cref{subsec:futurework}).

%% file: figures/fig_code_distr.tex
\definecolor{Regressed}{HTML}{C06969}
\definecolor{Misaligned}{HTML}{D9798C}
\definecolor{Uncorrected}{HTML}{E7BCBC}
\definecolor{Aligned}{HTML}{5CA25C}
\definecolor{Enlightened}{HTML}{95DA95}
\definecolor{Confirmed}{HTML}{B9DFB9}
\definecolor{Undermined}{HTML}{9D9D9D}
\definecolor{Uninformed}{HTML}{BABABA}
\definecolor{Uncertain}{HTML}{D3D3D3}

\begin{figure*}[t]
\begin{subfigure}[t]{0.5\textwidth}
\centering
\raisebox{1.\height}{
\fontsize{7}{8}\selectfont
\begin{tabular}{l|p{0.22\linewidth}|p{0.22\linewidth}|p{0.22\linewidth}}
\toprule
\textbf{Pre / Post} & \textbf{Flawed} & \textbf{No} & \textbf{Accurate} \\
\midrule
\textbf{Flawed} & \cellcolor{Uncorrected}Uncorrected (+0) \newline (14.6\%) 
                & \cellcolor{Uncertain}Uncertain (+1) \newline (1.5\%) 
                & \cellcolor{Aligned}Aligned (+2) \newline (19.6\%) \\
\midrule
\textbf{No} & \cellcolor{Misaligned}Misaligned (-1) \newline (5.0\%) 
            & \cellcolor{Uninformed}Uninformed (+0) \newline (3.5\%) 
            & \cellcolor{Enlightened}Enlightened (+1) \newline (13.6\%) \\
\midrule
\textbf{Accurate} & \cellcolor{Regressed}Regressed (-2) \newline (1.0\%) 
                   & \cellcolor{Undermined}{Undermined} (-1) \newline (0\%) %
                   & \cellcolor{Confirmed}Confirmed (+0) \newline (41.2\%) \\
\bottomrule
\end{tabular}
}
\caption{Pre-post shifts of GenAI understanding.}
\label{subfig:code_distr_matrix}
\end{subfigure}%
\hfill
\begin{subfigure}[t]{0.46\textwidth}
\centering
\includegraphics[width=\linewidth]{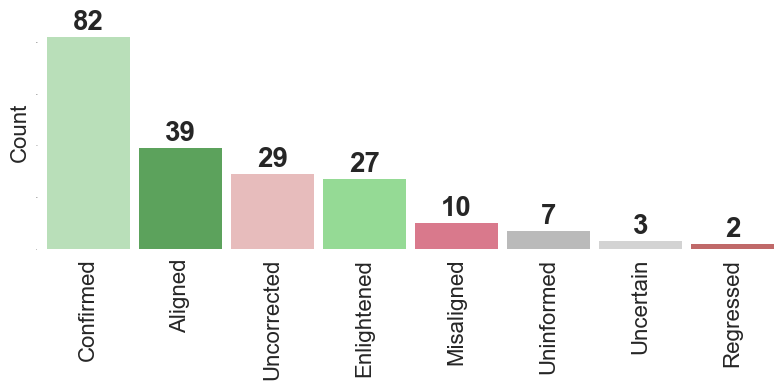}
\caption{Distribution of understanding shifts.}
\label{subfig:code_distr_bars}
\end{subfigure}

\caption{Pre- to post-game shifts in participants' flawed/ no/ accurate understandings of GenAI behaviors. Empty or unrelated responses are excluded.} %
\label{fig:code_distr}
\vspace{-10pt}
\end{figure*}

%% file: sections/discussion.tex
\section{Discussion and Future Work}
\label{sec:discuss}

\label{subsec:fast-ai-design}

\paragraphBold{Design for the fast progressing AIs.}

A recurring challenge during designing \sysname is keeping it relevant amid the rapid advancement of GenAI model capacity. %
In \sysname, we chose to emphasize \emph{constrained prompting} that reveals model failures with \emph{under-} or \emph{ambiguous} instructions -- issues unlikely to be resolved through incremental updates, since \textbf{underspecification is a fundamentally human trait}~\cite{Chenyang2025-qo,Ma2025-wf}.
This highlights a crucial design implication: GenAI literacy games -- and GenAI tools more broadly -- should not rely on specific model limitations. We should focus on more generalizable insights aligned with the model's growing ability to follow instructions. 
Looking ahead, we speculate that interventions on potential negative impact of AI usage would offer more lasting value. For example, future games could expose how user-granted access affects a model's invasiveness, or how degrees of personalization trade off helpfulness against risks of information leakage.

\paragraphBold{Designing for unconstrained GenAI by constraining hypothesis testing.}
As hypothesis testing is crucial in fostering critical reflection, a core challenge in designing reflective gameplay for GenAI systems is that the open-ended nature of prompting essentially means an unlimited hypothesis space.
We experimented with varying levels of prompting freedom, from freely writing any prompt to highly structured, fill-in-the-blank prompts. However, players either easily felt lost as they did not know where to start, or the prompts were too trivial for modern models.
In \sysname, we ultimately converged on a ``prompt reverse engineering'' task like \textit{Drawful} \cite{drawful}, but added a key constraint to \emph{make the prompt as short as possible}. 
This design preserved \rplayele{hypothesis-testing} and \rplayele{self-explanation} dynamics while encouraging players to reflect on models' default behaviors in bounded prompt space. 
Our results show that constrained prompts still foster thought-provoking hypotheses testing and refinement of mental models.
This lesson echoes prior GenAI literacy interventions, where effective scaffolding could require restricting human-GenAI interactions \cite{Ma2025-wf,Ma2024-gy}.

\paragraphBold{Limitations and Future Work}
\label{subsec:limit}
\label{subsec:futurework}

In our study, we have seen promising patterns suggesting shifts toward a more accurate understanding from pre- to post-game, as well as differences among self-rated experts and novices. %
We contribute to formulating an important aspect of GenAI literacy: critical reflections on GenAI behaviors. However, our results are constrained by the small sample size and study setup. 
\cmedit{We cannot disentangle the contribution of game mechanics from content exposure without a control condition, and the short interval between assessments may introduce potential testing effects.}
Future work could extend our preliminary findings to develop more validated and scalable constructs and test them with more diverse populations, such as children or multilingual settings.

Both the game mechanism and the backend GenAI models can also be updated to deepen and diversify reflections.
We may adopt more deliberate game mechanisms like AI voting, human drawing, prompt iteration, adversarial mechanics (e.g., creating challenging prompts/images for others, riffing on others' prompts), or team-based play. 
In our game, we instantiate GenAI with the specific \texttt{gpt-image-1}, but our design lessons and mechanisms can be generalized to other GenAIs. Enabling players to use and choose models with different capacities, such as DALL-E, may further open doors for in-depth understandings of different GenAI's limitations and strengths. %
\cmedit{Beyond education, scaled-up \sysname could support ML research by generating rich prompts and human judgments, which could complement existing datasets \cite{quickdraw} and support tasks like perceptual similarity benchmarking \cite{Trinh2025-eu}.}